\documentclass[12pt]{article}
\usepackage{axodraw}

\newcommand{\bea}{\begin{eqnarray}}
\newcommand{\eea}{\end{eqnarray}}

\title{{\small\hfill IMSc/2004/10/35}\\
\textbf{Infrared behaviour of massless QED in space-time dimensions
$2 < d < 4$}}

\author{Indrajit Mitra$^{a,b}$\footnote{indra@theory.saha.ernet.in},
Raghunath Ratabole$^{a}$\footnote{raghu@imsc.res.in}~ and
H. S. Sharatchandra$^{a}$\footnote{sharat@imsc.res.in} \\\\
$^a$ The Institute of Mathematical Sciences, C.I.T. Campus, Taramani P.O.,\\
Chennai 600113, India \\
$^b$ Theory Group, Saha Institute of Nuclear Physics, 1/AF Bidhan-Nagar,\\
Kolkata 700064, India}
\date{}
\begin{document}
\maketitle
\begin{abstract}

We show that the logarithmic infrared divergences in electron 
self-energy and vertex function of massless QED in 2+1 dimensions
can be removed at all orders of $1/N$ by an appropriate choice 
of a non-local gauge. Thus the infrared behaviour given by the 
leading order in $1/N$ is not modified by higher order corrections. 
Our analysis gives a computational scheme for the Amati-Testa model,
resulting in a non-trivial conformal invariant field theory
for all space-time dimensions $2 < d < 4$.

\end{abstract}
\newpage

Massless QED in 2+1 dimensions is of 
interest for various reasons. It provides a 
theoretical laboratory for studying the infrared (iR)
divergences
of perturbation theory \cite{jt, ap, t, ah, radu, abkw, anw, n, aw}
and chiral symmetry breaking \cite{abkw, anw, n, aw, mavro, bkp, fadm}. It also arises 
naturally in several theories of high temperature 
superconductivity \cite{rw, ghr, ftv}.
Also remarkably the theory is not simply super-renormalizable, it is 
ultraviolet (uV) finite.
For a Green function with
$F$ ($B$) number of external fermion (boson) lines, the 
superficial degree of divergence is
$\delta (F,B)=4-(3/2)F-B-L$, 
where $L$ is the number of loops. Consider
the possible uV divergent diagrams: 

1. One-loop fermion self-energy $\Sigma (p)$ has $\delta=0$. But the mass
renormalization is absent as a consequence of chiral symmetry.
Therefore, the contribution is uV finite.

2. One-loop vacuum polarization $\Pi_{\mu\nu}(q)$  has $\delta=1$. But 
gauge invariance requires that it has the form 
\bea
\Pi_{\mu\nu}(q)=(q^2\delta_{\mu\nu}-q_\mu q_\nu)\Pi(q^2).  \label{pi}
\eea
(We consider Euclidean Green functions throughout this paper.)
As two powers of the photon momenta are pulled out, 
$\Pi (q^2)$  effectively has $\delta=-1$ and therefore
it is uV finite.

3. Two-loop vacuum polarization has $\delta=0$. It is also uV
finite due to gauge invariance.

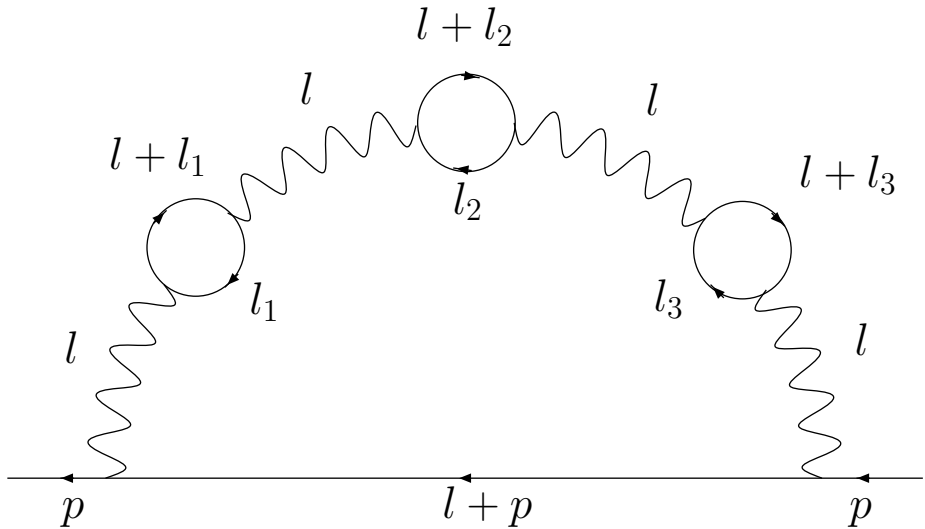
\begin{figure}[!b]
\begin{center}
\begin{picture}(351,198) (120,-181)
\SetWidth{0.5}
\ArrowLine(465,-162)(120,-162)
\CArc(293,-28)(18.38,135,495)
\CArc(191,-75)(18.38,135,495)
\CArc(397,-76)(18.38,135,495)
\ArrowLine(291,-10)(298,-11)
\ArrowLine(295,-46)(288,-45)
\ArrowLine(390,-94)(385,-89)
\ArrowLine(207,-85)(202,-90)
\PhotonArc(282.33,-175.33)(150.1,47.88,78.99){-7.5}{4.5}
\PhotonArc(303.94,-159.79)(123.08,-1.03,33.98){-7.5}{4.5}
\Text(212,-103)[lb]{\Large{{$l_1$}}}
\Text(365,-102)[lb]{\Large{{$l_3$}}}
\Text(231,-21)[lb]{\Large{{$l$}}}
\Text(362,-25)[lb]{\Large{{$l$}}}
\Text(441,-116)[lb]{\Large{{$l$}}}
\Text(159,-48)[lb]{\Large{{$l+l_1$}}}
\Text(420,-55)[lb]{\Large{{$l+l_3$}}}
\Text(275,1)[lb]{\Large{{$l+l_2$}}}
\Text(289,-65)[lb]{\Large{{$l_2$}}}
\Text(142,-119)[lb]{\Large{{$l$}}}
\PhotonArc(289.93,-156.16)(128.15,97.14,132.71){-7.5}{4.5}
\PhotonArc(298.78,-164.91)(141.81,147.63,178.82){-7.5}{4.5}
\ArrowLine(408,-62)(413,-67)
\ArrowLine(175,-66)(180,-61)
\ArrowLine(445,-162)(440,-162)
\ArrowLine(145,-162)(140,-162)
\Text(141,-181)[lb]{\Large{{$p$}}}
\Text(285,-181)[lb]{\Large{{$l+p$}}}
\Text(439,-181)[lb]{\Large{{$p$}}}
\end{picture}
\caption[]{\sf If the $d^3l$ integration is performed first,
the Feynman integral diverges even for hard external momentum.
\label{f:problem}}
\end{center}
\end{figure}

The same power counting shows that the iR divergences become 
increasingly worse with the number of loops. 
The iR superficial degree of divergence is given by
$\Delta=-\delta$. In fact, there is a more severe type of iR
divergence in perturbation theory.  Self-energy 
insertions on any internal line of a loop 
give rise to infrared divergent contributions even 
for hard external momenta \cite{tH}. For the example shown in
Fig.\ \ref{f:problem}, let us perform the
$d^3l$ integration first. Clearly, the integrand will contain 
more and more
factors of $1/l^2$ with increasing number of self-energy insertions
on the photon line.
There is a similar problem with self-energy insertions on any internal
fermion line of a loop. As a result, perturbation theory does not exist.
%

A resummation of perturbation theory using 
$1/N$ expansion dramatically alters the situation \cite{ap}.
The Lagrangian is
\bea
{\cal L}=-\frac{1}{4}(\partial_\mu A_\nu-\partial_\nu A_\mu)^2
          +\sum_i \bar\psi^i(i\rlap/\partial-e\rlap/A)\psi^i
                                           \label{L}
\eea
where $\psi^i ~( i=1,\cdots N)$ 
are $N$ species of charged 
four-component spinors,
all massless. 
Enforcing $U(2N)$ invariance \cite{abkw} 
ensures that fermion 
mass is not generated to any order in
perturbation theory. The charge $e$ has an engineering
dimension 1/2. We take the large $N$ limit with
$Ne^2$ fixed.
Then to the leading order in $1/N$, the one-loop vacuum polarization 
$\Pi^{(1)}(q^2)$ 
has to be included with the free photon
propagator. Due to the masslessness of the fermion, $\Pi^{(1)}(q^2)$
is singular at $q=0$ \cite{jt, ap}:
\bea
\Pi^{(1)}(q^2)=\frac{\mu}{q},~~~\mu=\frac{Ne^2}{8}.          \label{Pi1}
\eea
With the conventional gauge choice,
\bea
D_{\mu\nu}(q)=\frac{ \delta_{\mu\nu}-q_\mu q_\nu/q^2}{q^2+\mu q}
               -(\alpha-1)\frac{q_\mu q_\nu}{q^4}.           \label{Dmunu}
\eea
This changes the infrared  
behaviour of the photon propagator from being 
inversely quadratic to inversely linear in momentum. 

We consider a rearranged 
perturbation theory, with this as the ``free"
photon propagator, but otherwise the usual  fermion
propagator and vertex. The only 
difference with the usual perturbation theory
is that the {\it one-loop} vacuum polarization
contributions are not to be included in the new 
diagrams. Now the ultraviolet
divergences are absent in any order as before, as the 
free photon propagator is as usual inversely quadratic 
for large momenta.
On the other hand the infrared behaviour is now very different.
The iR superficial degree of divergence is now
\bea
\Delta(F,B)=B+F-3.                                 \label{super}
\eea
It is to be observed that
this is independent of the number of loops and
depends only on the number of external lines.
This is analogous to the uV degree of
divergence of a renormalizable theory. 
For non-exceptional Euclidean momenta $\{q\}$ which 
go to zero uniformly like 
\bea
q=\rho Q,                                              \label{rho}
\eea
the Green function
is singular as $\rho^{-\Delta}$. 
In effect the scale dimension of
photon has changed from the canonical 1/2 to 1, while
that of the fermion remains at the canonical value 1.
This infrared limit corresponds to an infrared stable fixed point \cite{abkw}.

Subintegrations can spoil the elegant picture of the iR
behaviour described above \cite{jt, ap, t, ah}. 
The danger is from subdiagrams
with $\Delta=0$ which can generate a logarithmic singularity in momenta
external to this subdiagram. Thus powers of logarithms arise from
various subdiagrams. These logs can shift the infrared 
behaviour away from that given by the naive fixed point described 
earlier.

The fermion self-energy naively has $\Delta=-1$. 
However, the absence of self-mass
correction makes the effective $\Delta$ zero. Indeed
an explicit calculation of $\Sigma(p)$ in one loop gives
a log correction \cite{jt, t, n}. 
See Ref.\ \cite{ftv} for computation of the anomalous dimension
of the fermion, suggesting that the canonical value is wrong.
This further casts doubt on the
relevance of the naive iR fixed point.
The vertex correction also has $\Delta=0$, and gives 
rise to logs in an arbitrary gauge.

However, the fermion self-energy depends on the gauge chosen, 
and we have to address the gauge invariant 
Green functions to unambiguously describe the
iR behaviour. The simplest such objects are the
Green functions involving only photons. As a consequence
of Ward identities, the photon fields appear only in the
field strength combination, and the Green functions
are gauge invariant. Now there are
sufficient indications \cite{ah, abkw, n, jsb} 
in the two-loop order that 
the logs from fermion self-energy and vertex corrections cancel
as a consequence of gauge invariance. It
is also conjectured that such cancellations 
take place at all orders \cite{ah, abkw}. But an explicit 
demonstration is lacking.

The situation is very similar to the uV logs
in QED in 3+1 dimensions. There are log divergences in
one-loop fermion self-energy and vertex 
corrections. When these are plugged into
a two-loop calculation of the vacuum polarization,
we expect two powers of log, as the vacuum polarization itself
has effective $\delta=0$. But an explicit calculation \cite{iz} shows
a cancellation of the squares of logarithms between the fermion
self-energy and vertex correction diagrams, so that
only one power of log results. Johnson, Willey and Baker \cite{jwb} 
have shown such a cancellation to all orders. Their 
strategy is to prove that a gauge choice exists in which 
the fermion self-energy and vertex corrections are
free of log divergence. 
We will adopt this approach here. The demonstration is,
however, much simpler in 2+1 dimensions.

We may expect that (with a specific choice of the
gauge), if log corrections in fermion self-energy and
vertex corrections are absent, 
such insertions into  other Green functions do not
lead to log corrections.  Then the
simple picture of iR behaviour is true. 

We should not expect the absence of log corrections
in higher orders even for a particular choice of the 
gauge parameter $\alpha$ in Eq.\ (\ref{Dmunu}).
The reason is that the $\alpha$-dependent part
is inversely quadratic in momentum and cannot
cancel the iR logs coming from the inversely linear part.
However we can choose the non-local Kondo-Nakatani gauge \cite{n, aw}
in which the photon propagator is
\bea
D_{\mu\nu}(q)=\frac{ \delta_{\mu\nu}-\xi q_\mu q_\nu/q^2}{q^2+\mu q}.
                                                      \label{Dnew}
\eea
Now the part dependent on the gauge parameter $\xi$ contributes to
the infrared divergences in the same way as the rest, and
there is hope that the log corrections vanish with a 
particular choice of the parameter. Indeed this 
has been checked in the lowest order
(see Ref.\  \cite{n} for the case of the fermion self-energy and 
Appendix \ref{app:vertex} for the vertex correction).

We demonstrate here that with an appropriate choice
of $\xi$ in each order of $1/N$ expansion,
the fermion self-energy and the
vertex corrections have no logarithmic infrared divergences, and
there are no log corrections to the leading $1/N$ infrared behaviour of any Green function.
Our proof is iterative. We presume this to hold to $O(N^{-n})$, and show that this
is then true to $O(N^{-n-1})$. (By a connected Green function of $O(N^{-n})$, 
we mean the following. Powers of $e$ in the corresponding tree diagram, if any,
are to be disregarded. Each remaining $e^2$ is to 
be replaced by $N^{-1}$. Finally,
each fermion loop contributes a factor of $N$.)

The Feynman integrals of a Green function $g(\{q\})$ have the following 
components depending on the external momenta $\{q\}$
and the loop momenta $\{l\}$:

(i) The photon propagators as in Eq.\ (\ref{Dnew}) with $q$ replaced with
    $\Sigma l+\Sigma q$, 
    where $\Sigma l$ ($\Sigma q$) denote appropriate linear combinations
    of the internal (external) momenta.

(ii) The fermion propagators $1/(\sum \rlap/l+\sum\rlap/q)$

(iii) Integration $\int d^3l/(2\pi)^3$ over each loop momentum.

The vertex factors do not depend on the momenta.
We now choose the Euclidean external momenta $\{q\}$ going to zero uniformly 
as in Eq.\ (\ref{rho}),
where $Q$ are of $O(1)$. We also take
$\{Q\}$ to be non-exceptional, i.e., no proper subset of $\{Q\}$
sums to zero.
Let us make a change in the variables 
of loop integrations: $l \rightarrow \rho L$.
Pulling out $\it one$ factor of $\rho$
from the denominator of each photon propagator and each
fermion propagator, we get
$g(\{q\})=\rho ^{-\Delta} G(\{Q\}, \rho)$.
Here $\Delta$ is simply the naive infrared degree of
divergence of the Green function, as given in Eq.\ (\ref{super}).
$G(\{Q\}, \rho)$ has the same 
expression as $g(\{Q\})$, apart from 
a modified photon propagator with a denominator
$\rho (\sum L+\sum Q)^2 + \mu |\sum L+\sum Q|$. 

Therefore, setting $\rho =0$ formally, we get an 
expression for $G(\{Q\}, \rho =0)$ which is exactly
that of QED, apart from a photon propagator with a denominator
$\mu |\sum L+\sum Q|$, i.e., inverse linear in momentum.
The other rules are unchanged  
(except that the one-loop vacuum polarization 
corrections are ignored). 

First note that as $Q$ are all of $O(1)$ and non-exceptional, 
$G(\{Q\},\rho=0)$ is iR finite (Appendix \ref{app:poggio}). 
However, $G(\{Q\},\rho\rightarrow 0)$
can have uV divergence,
with the uV superficial degree of divergence 
just the negative of that given in Eq.\ (\ref{super}).
Thus the situation is as in a renormalizable theory,
with uV divergences only in self-energy and
vertex parts at all orders. (As a consequence of 
gauge invariance, the photon self-energy corrections are
anyway free of uV divergences even now. Also, overlapping
divergences within the photon self-energy corrections can
be handled in the usual way.)
In effect, we have mapped the iR divergence of $g(\{\rho Q\})$
for $\rho\rightarrow 0$ to the uV divergence of 
$G(\{Q\},\rho\rightarrow 0)$. The cut-off for the uV divergence
is provided by $1/\rho$.
We have presumed that by a choice of the 
gauge parameter, the fermion self-energy and
vertex corrections of lower orders contained in 
our diagrams have been rendered finite. Then 
$G(\{Q\},\rho\rightarrow 0)$
is finite except when it is a fermion self-energy
or vertex correction.

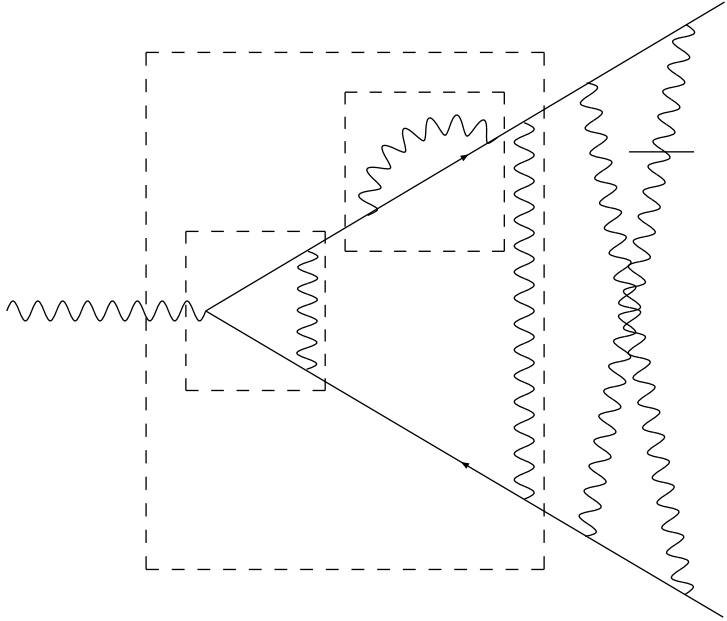
\begin{figure}
\vspace{-60pt}\hfill\\
\begin{picture}(480,390) (90,-135)
\SetOffset(120,0)
\SetWidth{1.0}
\SetScale{0.5}
\Photon(90,133)(240,133){7.5}{8}
\Photon(317,178)(316,89){7.5}{5.5}
\Photon(480,275)(480,-9){7.5}{14.5}
\PhotonArc(426.5,208.75)(64.61,58.77,183.33){-7.5}{7.5}
\DashLine(345,298)(465,298){10}
\DashLine(465,298)(465,178){10}
\DashLine(465,178)(345,178){10}
\DashLine(345,178)(345,298){10}
\DashLine(225,193)(225,73){10}
\DashLine(225,73)(330,73){10}
\DashLine(330,73)(330,193){10}
\DashLine(330,193)(225,193){10}
\DashLine(495,328)(495,-62){10}
\DashLine(495,-62)(195,-62){10}
\DashLine(195,-62)(195,328){10}
\DashLine(195,328)(495,328){10}
\ArrowLine(240,133)(631,366)
\ArrowLine(630,-98)(240,133)
\Photon(527,305)(601,-81){7.5}{20.5}
\Photon(602,349)(526,-37){7.5}{20.5}
\Line(559,253)(608,253)
\end{picture}
\vspace{-80pt}\hfill\\
\caption[]{\sf A contribution to the coefficient of $\ln\rho$.
\label{f:skeleton}}
\end{figure}

Now we concentrate on the case of the 
vertex correction. We enclose the self-energy
and vertex parts within boxes as in Fig.\ \ref{f:skeleton}. 
Since these (lower order)
contributions are presumed iR finite by appropriate choice
of the gauge parameter, we may set $\rho=0$
for such boxes. We may shrink the boxes to points and get the 
skeleton diagram, which has uV superficial degree of divergence
$\delta=0$. These points are
assigned $\rho$ independent factors. Since all proper
subdiagrams of the skeleton diagram have $\delta <0$, there
is just a $\ln\rho$ divergence and not a higher power of
logarithm.

We now show how to calculate the
coefficient of the part which diverges as $\ln \rho$. Let us apply  
the operation
$[\rho (d/d\rho)]|_{\rho=0}$ 
on the skeleton diagram. The operation $d/d\rho$ 
modifies the denominator of each photon 
propagator of the skeleton, one at a time, to
$(\rho|\Sigma L+\Sigma Q| +\mu)^2$.
We denote this modified photon
propagator by a cut (see Fig.\ \ref{f:skeleton}).
In order to be able to take the limit
$\rho\rightarrow 0$, we scale the loop variables back to $l$, i.e.,
replace $L$ with $l/\rho$. As $\delta=0$, the net effect of
$[\rho (d/d\rho)]|_{\rho=0}$ is 
the original expression for the skeleton evaluated at zero external 
momenta, except that
the denominator of the cut photon propagator is modified
to $(|\Sigma l| + \mu )^2$. (The denominators of the other
photon propagators are $(\Sigma l)^2+\mu |\Sigma l|$.)
First, note that the photon propagators are inverse quadratic in momentum
for large momenta just like the usual propagator,
and hence our expression is uV finite.
Secondly, note that $\mu$ provides an iR cutoff to the cut photon
propagator, resulting in $\Delta=-1$.
This again confirms that there is just a $\ln\rho$ divergence, 
and explicitly determines the numerical coefficient of $N^{-n-1}$.
In Appendix \ref{app:vertex}, this procedure for 
the extraction of the
$\ln\rho$ term is illustrated by working out the case of the $O(1/N)$ vertex.

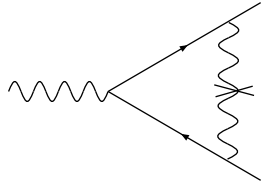
\begin{figure}
\begin{picture}(190,134) (45,-143)
\SetOffset(200,0)
\SetWidth{1.0}
\SetScale{0.5}
\Photon(45,-76)(120,-76){7.5}{4}
\Photon(211,-24)(210,-127){7.5}{4.5}
\ArrowLine(120,-76)(235,-9)
\ArrowLine(235,-143)(120,-76)
\Line(200,-71)(230,-79)
\Line(201,-80)(229,-72)
\end{picture}
\vspace{-80pt}\hfill\\
\caption[]{\sf Contribution from adjustment of the gauge parameter $\xi$
which cancels the logarithmic divergence from higher order vertex.
\label{f:counterterm}}
\end{figure}

We now adjust $\xi$ at $O(N^{-n})$ so that the contribution
from the $O(1/N)$ vertex
cancels this log divergence (Fig.\ \ref{f:counterterm}).
Then the vertex is iR finite to $O(N^{-n-1})$.
As a consequence of the Ward identity, $Z_1=Z_2$,
the fermion self-energy will also be finite to this order, thus
completing the proof.

%

Here it is to
be emphasized that there is a major difference with
respect to the case of uV divergences.
Once counterterms are in place to remove the uV 
divergences, all subintegrations have $\delta <0$,
and the loop integrations can be carried out in 
any order with a unique result (i.e., absolute convergence). But in the present
situation the self-energy insertions on a propagator
have to be evaluated first, and fed into the other 
loop integrations (i.e., the integrals are only conditionally convergent). 
(See Appendix \ref{app:poggio}.)
As an example, consider Fig.\ \ref{f:problem} with higher order vacuum
polarization insertions.
We do not have the luxury of carrying
out the $d^3l$ integration first.
When the vacuum polarization insertions are evaluated, each of them
is proportional to $l$, while each photon propagator is
proportional to $1/l$. So the problem referred to in Fig.\ \ref{f:problem}
is now absent.

%

We have shown that with a specific choice of the gauge parameter
in Eq.\ (\ref{Dnew})
(to each order in $1/N$), all Green functions $G(\{Q\},\rho\rightarrow 0)$
are finite. 
These are the Green functions of a theory which uses 
the photon propagator
$(\delta_{\mu\nu}-\xi q_\mu q_\nu/q^2)/(\mu q)$
and the usual QED rules otherwise. The one-loop vacuum 
polarization is not to be included in corrections to this 
photon propagator. Indeed the photon propagator is just
this one-loop contribution. Thus this corresponds to
simply the functional integral
\bea
\int \prod_i {\cal D}\psi^i{\cal D}\bar\psi^i{\cal D}A_\mu \exp[i\int
          \sum_i\bar\psi^i(i\rlap/\partial-e\rlap/A)\psi^i].      
	                                                    \label{funcint}
\eea
Integrating over 
$A_\mu$, we have
\bea
\int\prod_i {\cal D}\psi^i{\cal D}\bar\psi^i \prod_x 
\delta(\sum_i \bar\psi^i\gamma^\mu \psi^i)
          \exp[i\int \sum_i \bar\psi^i i\rlap/\partial \psi^i].
\eea
This is the Amati-Testa model \cite{at}. 
If we scale the photon field
$A_\mu \rightarrow  (1/e) A_\mu$ in Eq.\ (\ref{L}), the Green functions
are formally of the QED theory 
\bea
{\cal L}=-\frac{1}{4e^2}(\partial_\mu A_\nu-\partial_\nu A_\mu)^2
          +\sum_i \bar\psi^i(i\rlap/\partial-\rlap/A)\psi^i
\eea
in the limit $e \rightarrow \infty$.
Thus $1/e$ plays the role of $\rho$ in our scaling Eq.\ (\ref{rho}).

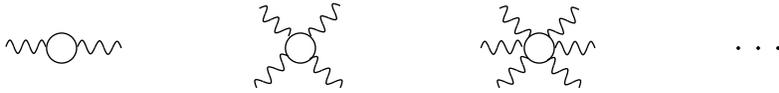
\begin{figure}
\vspace{80pt}\hfill\\
\begin{picture}(583,67) (33,-233)
\SetOffset(100,0)
\SetWidth{1.0}
\SetScale{0.5}
\CArc(75,-203)(11.31,135,495)
\CArc(255,-203)(11.31,135,495)
\CArc(435,-203)(11.31,135,495)
\Vertex(585,-203){1.41}
\Vertex(600,-203){1.41}
\Vertex(615,-203){1.41}
\Photon(87,-202)(120,-203){5}{2.5}
\Photon(405,-173)(427,-195){5}{2.5}
\Photon(391,-203)(422,-202){5}{2.5}
\Photon(443,-195)(465,-174){5}{2.5}
\Photon(427,-211)(405,-233){-5}{2.5}
\Photon(222,-233)(246,-209){5}{2.5}
\Photon(262,-211)(285,-233){5}{2.5}
\Photon(447,-203)(477,-203){5}{2.5}
\Photon(441,-212)(465,-233){5}{2.5}
\Photon(224,-173)(246,-194){5}{2.5}
\Photon(263,-194)(285,-171){5}{2.5}
\Photon(33,-202)(63,-202){5}{2.5}
\end{picture}
\vspace{-120pt}\hfill\\
\caption[]{\sf The terms in the effective Lagrangian for a class
of conformal invariant field theories in dimensions $2 < d <4$.
\label{f:conformal}}
\end{figure}

The coupling constant $e$ provides the interpolation of the 
Green functions from the uV to the iR behaviour: 
$e\rightarrow 0$ gives the free theory of photons and
fermions as expected from the asymptotic freedom, while
$e\rightarrow \infty$ gives the infrared limit of the
Green functions. This latter limit is a non-trivial 
scale and conformal invariant 
theory with non-canonical scaling dimension for the photon.
In fact we have a line of fixed points labelled by $N$
(regarded as a continuous parameter). Integrating over
$\psi^i$, $\bar \psi^i$ in Eq.\ (\ref{funcint}), we get
\bea
\int{\cal D}A_\mu\exp[N {\rm Tr}\ln(1-\frac{1}{i\rlap/\partial}\rlap/A)].
\eea
We see that $N$ plays the role of $1/\hbar$, 
and $N \rightarrow \infty$ can be interpreted 
as the semi-classical limit of
the theory . The
Green functions in this limit are obtained as 
follows: Consider only the tree diagrams built from the theory
\bea
{\cal L}=-\sum^\infty_{2,4,\cdots}\frac{1}{n}{\rm Tr}(\frac{1}{i\rlap/\partial}\rlap/A)^n.
\eea
(See Fig.\ \ref{f:conformal}.) The $n=2$ term gives the inverse of the
propagator, while the other terms give the effective vertices.
Thus we have an explicit non-trivial conformal invariant theory in
2+1 dimensions. 

The other conformal field theories corresponding to other $N< \infty$
are obtained by using the propagator and effective vertices of the
Lagrangian given in Fig.\ \ref{f:conformal}, and
including the loop corrections with a factor of $1/N$ for every loop.
Thus $1/N$ provides a marginal operator
that takes us from the simplest conformal field theory to the entire class 
labelled by $N$. It is of great interest to construct these
conformal field theories explicitly.

It is very interesting that the analysis of this paper is valid
for all (Euclidean) dimensions $2 < d < 4$. 
The theory is uV finite in this range:
$\delta=4-(4-d)L-(3/2)F-B$. From the one-loop vacuum polarization, we now
get $\Pi^{(1)}(q^2)\sim q^{d-4}$ for $q\rightarrow 0$. Thus the iR
behaviour of our 
photon propagator will be $q^{2-d}$. Then the
iR superficial degree of divergence is given by
\bea
\Delta=B+\frac{d-1}{2}F -d.
\eea
For the fermion self-energy, $\Delta=-1$, which in the absence of 
self-mass gives iR log as in $d=3$. Same is the case for 
the vertex
function which has $\Delta=0$. Again we can choose a gauge such that
these are iR finite and there are no log corrections. The iR limit
is a conformal field theory where the photon has non-canonical scaling 
dimension one for the entire range of $d$, in contrast to the
engineering dimension $(d-2)/2$.

%

In this paper, we demonstrated that the infrared behaviour given by
the leading order in $1/N$ is not modified by logarithmic corrections 
in higher orders. Our technique gives finite Green functions for the 
Amati-Testa model, and results in a non-trivial conformal field theory
for each $N$ in all space-time dimensions $2 < d < 4$. 

We have used the technique of choosing the value of the gauge parameter $\xi$
such that the logarithmic infrared divergence in the electron self-energy
and the vertex function is removed to all orders. This choice of $\xi$ will 
therefore simplify the
calculation of any gauge-invariant quantity (not necessarily a Green function)
in which these insertions occur. Examples of such gauge-invariant quantities
are gauge-invariant composite operators, related to the response functions
of condensed matter systems \cite{fpst}. Moreover, since for this value
of $\xi$ we know the infrared behaviour of a Green function to all orders,
the determination of the infrared behaviour of a gauge-variant Green function
to all orders for {\it any} value of $\xi$ becomes possible. This has implications for
the anomalous dimension of the dressed gauge-invariant fermion. These calculations
will be presented elsewhere.

\section*{Acknowledgment}

The LaTeX codes for the figures in this paper were generated primarily using
JaxoDraw \cite{jaxo}.

\appendix

\leftline{\null\hrulefill\null}\nopagebreak
\section*{Appendices}

\section{Logarithmically divergent part in vertex correction
to O(1/N)}\label{app:vertex}
The $O(1/N)$ correction to the QED vertex, with incoming
fermion momentum $p$ and outgoing fermion momentum $p+q$, is given by
\bea
e\Gamma_\mu^{(1)}(p+q,p)=\int\frac{d^3l}{(2\pi)^3}\,\,e\gamma_\sigma
         \frac{1}{\rlap/p +\rlap/q+\rlap/l}\, e\gamma_\mu
	 \frac{1}{\rlap/p+\rlap/l}\,e\gamma_\rho
	 \frac{\delta_{\sigma\rho}-\xi l_\sigma l_\rho/l^2}{l^2+\mu l}.
\eea
(We use the Euclidean space Feynman rules and gamma matrix algebra of
Ref.\ \cite{iz}.) 
To evaluate these integrals with the modified photon propagator,
we may use the spectral representation
\bea
\frac{1}{l^2+\mu l}=\frac{2\mu}{\pi}\int_0^\infty dM \frac{1}{(M^2+\mu^2)
                    (l^2+M^2)}                             \label{spec}
\eea
(see Ref.\ \cite{jt}), and
\bea
\frac{1}{l^2(l^2+\mu l)}=\frac{2\mu}{\pi}\int_0^\infty dM \frac{1}{M^2(M^2+\mu^2)}
                          \Bigg(\frac{1}{l^2}-\frac{1}{l^2+M^2}\Bigg) 
\eea
(as obtained from Eq.\ (\ref{spec})).
However here we consider only the iR behaviour.
Choose $p_\mu=\rho P_\mu$ and $q_\mu=\rho Q_\mu$. Also let 
$l_\mu=\rho L_\mu$. Then,
\bea
\Gamma_\mu^{(1)}(\rho(P+Q),\rho P)=e^2\int\frac{d^3L}{(2\pi)^3}\,\,\gamma_\sigma
         \frac{1}{\rlap/P +\rlap/Q+\rlap/L}\, \gamma_\mu
	 \frac{1}{\rlap/P+\rlap/L}\,\gamma_\rho
         \frac{\delta_{\sigma\rho}-\xi L_\sigma L_\rho/L^2}{\rho L^2+\mu L}.   \label{gama}
\eea
For $\rho\rightarrow 0$, this is iR finite but logarithmically uV divergent. Let the
divergent part be $C\ln\rho$. The coefficient $C$ is  obtained by the action of
$[\rho(d/d\rho)]_{\rho=0}$ on the R.H.S.of Eq.\ (\ref{gama}). Thus,
\bea
C=-e^2\Bigg[\rho\int\frac{d^3L}{(2\pi)^3}\,\,\gamma_\sigma
         \frac{1}{\rlap/P +\rlap/Q+\rlap/L}\, \gamma_\mu
	 \frac{1}{\rlap/P+\rlap/L}\,\gamma_\rho
         \frac{\delta_{\sigma\rho}-\xi L_\sigma L_\rho/L^2}{(\rho L+\mu)^2}\Bigg]_{\rho=0}.
\eea
The problem with this form is that we have $\rho$ times an integral which diverges 
at $\rho=0$. To get around
this problem, let us replace $L$ by $l/\rho$ in the integral. Then setting $\rho=0$, we obtain
\bea
C=-e^2\int\frac{d^3l}{(2\pi)^3}\,\,\gamma_\sigma \frac{\rlap/l}{l^2}
        \, \gamma_\mu \frac{\rlap/l}{l^2}\,\gamma_\rho
        \frac{\delta_{\sigma\rho}-\xi l_\sigma l_\rho/l^2}{(l+\mu)^2}.
\eea
The numerator of the integrand is $l^2(1-\xi)\gamma_\mu-2\rlap/l \,l_\mu$.
By symmetry, $\rlap/l\, l_\mu$ may be replaced with $(1/3)l^2\gamma_\mu$. 
Doing the angular integration, we arrive at
\bea
C&=&\frac{e^2}{2\pi^2}(\xi-\frac{1}{3})\gamma_\mu\int_0^\infty dl\,\frac{1}{(l+\mu)^2} \nonumber\\
 &=& \frac{4}{\pi^2 N}(\xi-\frac{1}{3})\gamma_\mu.			   
\eea
Thus, for the gauge choice $\xi=1/3$, there is no log
in the $O(1/N)$ vertex. As expected,this is
the same gauge in which the $O(1/N)$ self-energy is also 
free from logarithm \cite{n}. 
\section{Absence of infrared divergences for hard and 
non-exceptional Euclidean external momenta}\label{app:poggio}

Consider a Green function for which the external momenta are
Euclidean, non-exceptional and of $O(1)$, while the internal
photon lines are inversely linear in momentum. We follow the
standard power-counting arguments for Euclidean momenta 
\cite{jwb, pq}
to show that the Green function is free from iR divergences.

The hard momenta of the external lines flow through some of the
internal lines also, and 
all these hard internal lines must be connected
due to the non-exceptional nature of the external momenta.
Since a hard internal line does not contribute to the iR
degree of divergence,
it may be contracted to a point for the present purpose.
Thus we arrive at a reduced diagram in which all the external lines
are joined at a single point. Out of this single point, let $b$
soft internal boson lines and $f$ soft internal fermion lines come out and join
to a subdiagram $S$ consisting entirely of soft internal lines.

In addition to usual QED vertices, $S$ can also contain
composite vertices arising out of the contraction of hard
loops which are not connected to the hard external lines.
In general, such a composite vertex can have $m$ boson and $n$ 
fermion lines. Let the number of such a vertex in $S$ be $V_{mn}$. 

Therefore we have to calculate the iR
superficial degree of divergence of a diagram without any
external lines, which contains one composite vertex having $b$ boson and $f$
fermion lines, (say) $V$ elementary vertices, and also $V_{mn}$ composite
vertices with various values of $m$ and $n$. Taking $i_B$ and $i_F$ to be
the number of internal boson and fermion lines, we have
\bea
\Delta&=&i_B+i_F-3L\,,\\
L&=&i_B+i_F-(1+V+\sum_{m,n}V_{mn})+1\,,\\
2i_B&=&b+V+\sum_{m,n} mV_{mn}\,,\\
2i_F&=&f+2V+\sum_{m,n} nV_{mn}
\eea
leading to
\bea
\Delta=-b-f+\sum_{m,n} (3-m-n)V_{mn}\,.
\eea
It then appears that the cases where $(m,n)$ are $(2,0)$, $(0,2)$ and
$(1,2)$, can lead to iR divergence by making positive or logarithmic
contribution to $\Delta$. But these are 
precisely the self-energy and vertex
insertions explicitly addressed in this paper.
Thus, if the subintegration involved in
the composite vertex is performed first, the problem disappears, as 
will be explained now. The vertex correction is rendered finite by a choice of
non-local gauge. Chiral symmetry ensures that the fermion self-energy 
insertion is proportional to
$\rlap/p$, and the proportionality constant is finite in the same 
non-local gauge. Thus $\Delta$ is actually diminished 
(and not increased) by one due to
a fermion self-energy insertion. The same change in $\Delta$
happens for a photon self-energy
insertion.

\end{document}